\documentclass[aps,prb,twocolumn,superscriptaddress]{revtex4-1}

\usepackage{xspace}
\usepackage{color}
\usepackage{siunitx}
\usepackage[colorinlistoftodos,prependcaption,textsize=tiny]{todonotes}
\usepackage{amsmath}
\usepackage{amssymb}
\usepackage[pdftex, colorlinks=true, linkcolor=blue, citecolor=blue, urlcolor=blue]{hyperref}
\usepackage{graphicx}
\DeclareGraphicsExtensions{.png,.pdf}

\usepackage{xcolor}

\newcommand*{\ns}{\ensuremath{n_{\text{s}}}\xspace}
\newcommand*{\lambdaF}{\ensuremath{\lambda_\mathrm{F}\xspace}}

\newcommand*{\VSD}{\ensuremath{V_{\mathrm{SD}}}\xspace}
\newcommand*{\Vtip}{\ensuremath{V_{\mathrm{tip}}}\xspace}
\newcommand*{\Vcav}{\ensuremath{V_{\mathrm{cav}}}\xspace}
\newcommand*{\Vqpc}{\ensuremath{V_{\mathrm{qpc}}}\xspace}
\newcommand*{\EF}{\ensuremath{E_\mathrm{F}}\xspace}
\newcommand*{\Ecav}{\ensuremath{\Delta E_\mathrm{cav}}\xspace}
\newcommand*{\lc}{\ensuremath{l_\mathrm{c}}\xspace}
\newcommand*{\VG}[2]{\ensuremath{V_{\mathrm{#1}}^{\mathrm{#2}}}\xspace}

\newcommand*{\Acav}{\ensuremath{A_{\mathrm{cav}}}\xspace}
\newcommand*{\Ut}{\ensuremath{U_{\mathrm{t}}}\xspace}
\newcommand*{\ut}{\ensuremath{u_{\mathrm{t}}}\xspace}

\newcommand*{\tippot}{tip-induced potential\xspace}

\newcommand*{\ri}{(1)\xspace}
\newcommand*{\rii}{(2)\xspace}
\newcommand*{\riii}{(3)\xspace}
\newcommand*{\riv}{(4)\xspace}
\newcommand*{\rv}{(5)\xspace}

\newcommand{\zurich}{Solid State Physics Laboratory, ETH Zurich, 8093 Zurich, Switzerland}
\newcommand{\strasbourg}{Universit\'e de Strasbourg, CNRS, Institut de Physique et de Chimie des Mat{\'e}riaux de Strasbourg, UMR 7504, F-67000 Strasbourg, France}
\newcommand{\munich}{Technical University of Munich, Physics Department, James Franck Str, D-85748 Garching, Germany}

\begin{document}



\title{Scanning gate experiments: from strongly to weakly invasive probes}


\author{R. Steinacher}
\email[]{richard.steinacher@phys.eth.ch}
\affiliation{\zurich}

\author{C. P{\"o}ltl}
\affiliation{\strasbourg}

\author{T. Kr{\"a}henmann}
\affiliation{\zurich}
\author{A. Hofmann}
\affiliation{\zurich}
\author{C. Reichl}
\affiliation{\zurich}
\author{W. Zwerger}
\affiliation{\zurich}
\affiliation{\munich}
\author{W. Wegscheider}
\affiliation{\zurich}
\author{R. A. Jalabert}
\affiliation{\strasbourg}
\author{K. Ensslin}
\affiliation{\zurich}
\author{D. Weinmann}
\affiliation{\strasbourg}
\author{T. Ihn}
\affiliation{\zurich}

\date{\today}

\begin{abstract}
An open resonator fabricated in a two-dimensional electron gas is used to explore the transition from strongly invasive scanning gate microscopy to the perturbative regime of weak tip-induced potentials. With the help of numerical simulations that faithfully reproduce the main experimental findings, we quantify the extent of the perturbative regime in which the tip-induced conductance change is unambiguously determined by properties of the unperturbed system.
The correspondence between the experimental and numerical results is established by analyzing the characteristic length scale and the amplitude modulation of the conductance change. In the perturbative regime, the former is shown to assume a disorder-dependent maximum value, while the latter linearly increases with the strength of a weak tip potential.  

\end{abstract}


\maketitle


\section{Introduction}

The advent of scanning probe techniques in the 1980s \cite{Bhushan2010,Binnig1982,Binnig1986} opened the door for locally investigating electron transport in high-mobility two-dimensional electron gases (2DEGs) and in 2DEG-based nanostructures at cryogenic temperatures. 
After the pioneering work of Eriksson \textit{et al.}\ \cite{Eriksson1996,Eriksson1996a} introducing the scanning gate microscopy (SGM) technique, the discovery of branched electron flow in 2DEGs near a quantum point contact \cite{Topinka2001} has been a major achievement.

The experimental technique makes use of a tip-induced electrostatic potential in the 2DEG-plane characterized by an amplitude \Ut in energy, and a characteristic radius $w$ (width) in space.
The imaging mechanism typically employed in experiments uses a strongly invasive tip potential $\Ut>\EF$, significantly altering the electron flow in the system as compared to the unperturbed situation. Here, $E_\mathrm{F}$ is the Fermi-energy of the 2DEG.

Theoretical attempts to link tip-induced conductance changes to local properties of the unperturbed system \cite{Jalabert2010} assumed the use of a weakly invasive tip with $\Ut\ll \EF$. 
A connection with the local density of states (LDOS) could be established in the case of a quantum point contact (QPC) operating in the regime of conductance quantization,\cite{Gorini2013,Ly2017} while less direct relationships were produced beyond this particular situation.\cite{Pala2008,gasparian1996} 

Until now, experimental limitations did not allow experimentalists to perform measurements with $\Ut\ll \EF$.
The reason is that the scattering matrix elements of a weak tip, as well as the LDOS at the Fermi energy are very small in an extended two-dimensional system. 
A significant measurement signal occurs only  if either the LDOS is strongly concentrated in space (closed geometries), or if large tip-induced potentials ($\Ut>\EF$) are used, which have the undesired effect of strongly modifying the local electronic properties under investigation.

In recent publications we have demonstrated that the scanning gate technique loses the fine resolution achieved in open 2DEGs once closed geometries are investigated.\cite{kozikov2014locally, Steinacher2016} 
In this paper we aim at the intermediate regime between a closed and an open geometry, attempting to close the gap between the interpretations of the two extreme cases. 
We progressively concentrate the LDOS in a ballistic cavity, considered as an open resonator, in order to allow for studying the transition from weakly to strongly invasive measurements. 
At the same time the LDOS is extended over a region significantly larger than the characteristic width $w$ of the \tippot, allowing us to exploit the spatial resolution capabilities of the technique.

\begin{figure}[ht]
\centering
\includegraphics[width=\linewidth]{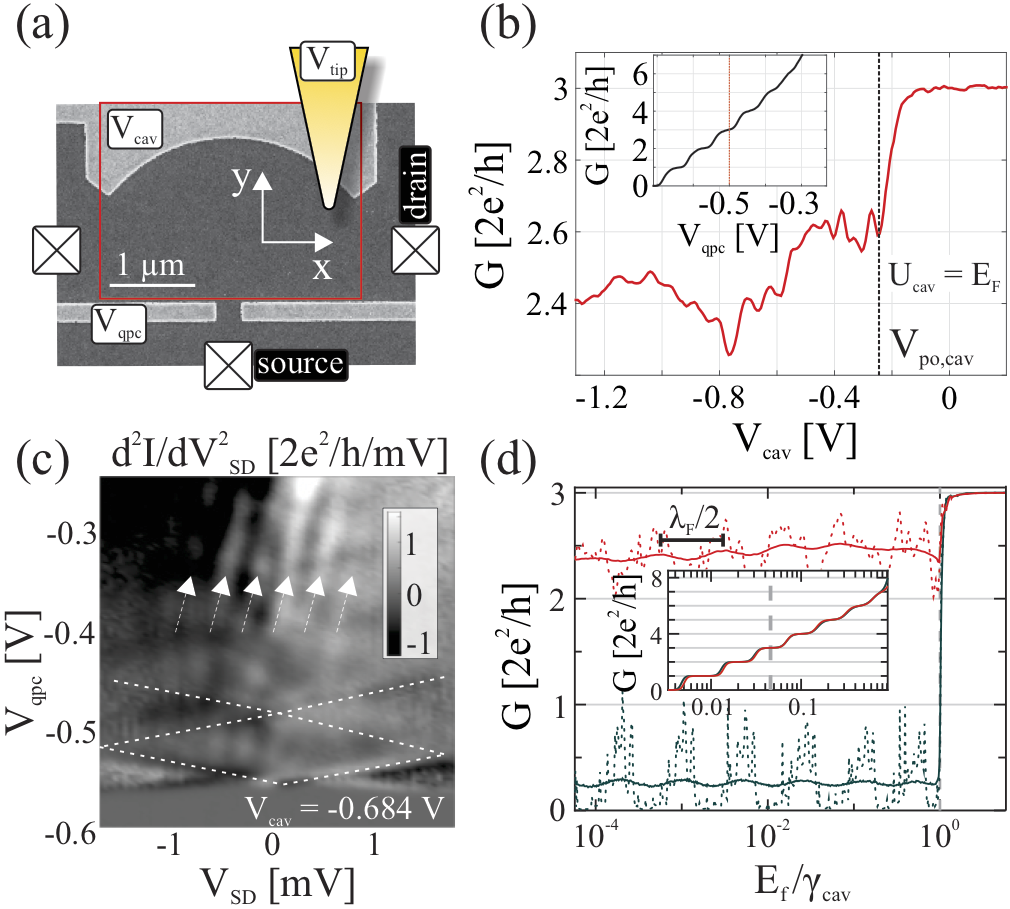}
\caption{(a) Scanning electron micrograph of the sample. 
Schottky gates of the quantum point contact and the cavity gate in light gray, GaAs surface in dark gray. 
Scan area outlined by the red rectangle. Black labels symbolize ohmic contacts. 
(b) Conductance $G$ as a function of the cavity-gate voltage $\Vcav$. 
The vertical dashed black line marks voltage of 2DEG depletion below the gate. 
Inset: $G$ as a function of $\Vqpc$ at $\Vcav=0$. The red line marks the working point $\Vqpc=\SI{-0.5}{\volt}$. 
(c) Numerical derivative $\mathrm{d}G/\mathrm{d}\VSD$ as function of \Vqpc and \VSD. Cavity modes highlighted with white arrows. 
(d) Numerical simulation of the conductance as a function of the height $\gamma_\mathrm{cav}$ of the cavity gate potential. 
Green-blue lines: no disorder. 
Red lines: realistic disorder. 
Dashed and solid lines: $T=0$ and $T=\SI{500}{\milli\kelvin}$, respectively. 
Inset: simulation of $G$ for $\gamma_\mathrm{cav}=0$ as a function of the QPC gate-potential $\gamma_\mathrm{QPC}$. 
The vertical line indicates the value used in the calculations as a function of $\gamma_\mathrm{cav}$.}
\label{fig1}
\end{figure}
\section{Sample and Experimental Setup}

The sample is based on a GaAs/AlGaAs heterostructure with a 2DEG \SI{90}{\nano\meter} below the surface. 
The electron gas has a density $\ns=\SI{1.5e11}{\centi\meter^{-2}}$ and an electron mobility $\mu = \SI{3.35e6}{\centi\meter^2/\volt\second}$ at the setup's base temperature of \SI{270}{\milli\kelvin}. 
Schottky gates (Ti/Au) defined by electron-beam lithography on top of a photolithography-defined mesa structure form the device in Fig.\ \ref{fig1}(a). 
The QPC visible at the bottom has a lithographic gap of \SI{300}{\nano\meter} and is controlled by the gate voltage  \Vqpc. 
The lower edge of the cavity gate (controlled by the gate voltage \Vcav) visible at the top describes a circular arc with its center in the QPC opening, a radius of \SI{2}{\micro\meter}, and an opening angle of \SI{90}{\degree}. 
The gates deplete the 2DEG at voltages below $\VG{depl,cav}{}=\SI{-0.25}{V}$.

The scanning tip is oriented normal to the surface of the structure with a voltage \Vtip applied relative to the 2DEG. 
We raster-scan the tip \SI{65}{\nano\meter} above the sample surface while measuring the two-terminal conductance of the device. 
We record maps $G(x,y)=I_\mathrm{SD}(x,y)/V_\mathrm{SD}$ of linear conductance versus tip-position $(x,y)$ by applying an alternating voltage $\VSD=\SI{100}{\micro\volt}\mathrm{rms}$ between source and drain (labeled in Fig.\ \ref{fig1}(a)) and measuring the alternating current $I_\mathrm{SD}$ with a home-built current--voltage converter and a commercial lock-in amplifier.

\section{Effect of the cavity in the absence of the scanning tip}
\label{section3}

We first characterize the conductance of the structure in the absence of the tip.
The inset of Fig.\ \ref{fig1}(b) shows the QPC conductance for $\Vcav=0$, as a function of the QPC gate voltage \Vqpc. 
A sequence of at least five conductance steps is seen quantized in units of $2e^2/h$. 
At $V_\mathrm{qpc}=\SI{-0.5}{V}$ the QPC transmits three spin-degenerate modes. 
This is the working point for the following measurements, indicated in the inset by the red vertical line.

At this fixed value of $V_\mathrm{qpc}$ the conductance of the structure depends strongly on the cavity voltage \Vcav, as seen in Fig.\ \ref{fig1}(b).
For $\Vcav \geq \VG{depl,cav}{}$, the depletion voltage below the cavity gate, only tiny changes in $G$ can be observed when $\Vcav$ is changed. 
Once the cavity gate depletes the 2DEG at $\Vcav = \VG{depl,cav}{}$ the conductance drops by roughly 15\,\%. At even smaller \Vcav the conductance shows mesoscopic modulations. 
They are caused by electron interference: the reflecting edge of the depletion region defining the cavity shifts linearly with \Vcav\cite{steinacher2015scanning} by an estimated amount of  \SI{350}{\nano\meter/\volt}. 
As a result of the shift, the phase and the amplitude of the reflected electron waves change with \Vcav.\cite{Katine1997,Roessler2015}
We estimate $\lambdaF/2=\SI{32}{\nano\meter}$, which is consistent with the number of oscillations along the bar of length $5\lambdaF/2$ in the figure, and confirm that the modulation of the conductance occurs on the scale of $\lambdaF/2$. 

Numerical simulations in the regime of quantum transport play a key role for the interpretation of scanning gate experiments.\cite{Poeltl2016} 
We model the experimental cavity with and without the action of a charged tip using the KWANT package\cite{Groth2014} which is based on the recursive Green function method\cite{Lee1981} for non-interacting electrons.  
Fig.\ \ref{fig1}(d) presents simulations showing the reflecting effect of the cavity potential (with an energy  $\gamma_\mathrm{cav}$ assumed to be uniform in the gated region) in the absence of the tip with and without disorder (red and green-blue lines, respectively), at zero temperature (dashed lines) and at $T=\SI{500}{\milli\kelvin}$ (solid lines). 
The QPC is set to operate on the third conductance plateau (see inset) as in the experiment. 
The disorder is modeled by remote impurities with the experimentally realized density and a distance of \SI{50}{\nano\meter} from the 2DEG. 
The resulting transport mean free path at the Fermi energy of $l_\mathrm{tr}=\SI{23.8}{\micro\meter}$ is very close to the experimental value ($\SI{22.4}{\micro\meter}$). 
The elastic mean free path is  $l=\SI{0.22}{\micro\meter}$.

Like in the experimental case, the cavity potential has a very weak effect on the conductance until it reaches a value comparable with the Fermi energy. 
Beyond this value, the cavity is defined by the depletion of the 2DEG and the disorder plays a crucial role. 
The clean cavity focuses the electrons back through the QPC, and thus the resulting conductance is very small. 
In the presence of disorder the conductance is only weakly suppressed by the reflecting effect of the cavity. 
Even for the weak disorder we consider here ($l_\mathrm{tr}$ is much larger than the cavity size, placing us in the ballistic regime), the conductance in the presence of the cavity is strongly enhanced as compared to the clean case. 
This feature demonstrates the importance of small-angle scattering and the dramatic effect of disorder, counteracting the reflecting trend of the cavity for arbitrary openings of the QPC. 
It allows us to relate the pinch-off step-height of the conductance with the disorder strength in the cavity.

\begin{figure*}[th]
\centering
\includegraphics[width=0.8\linewidth]{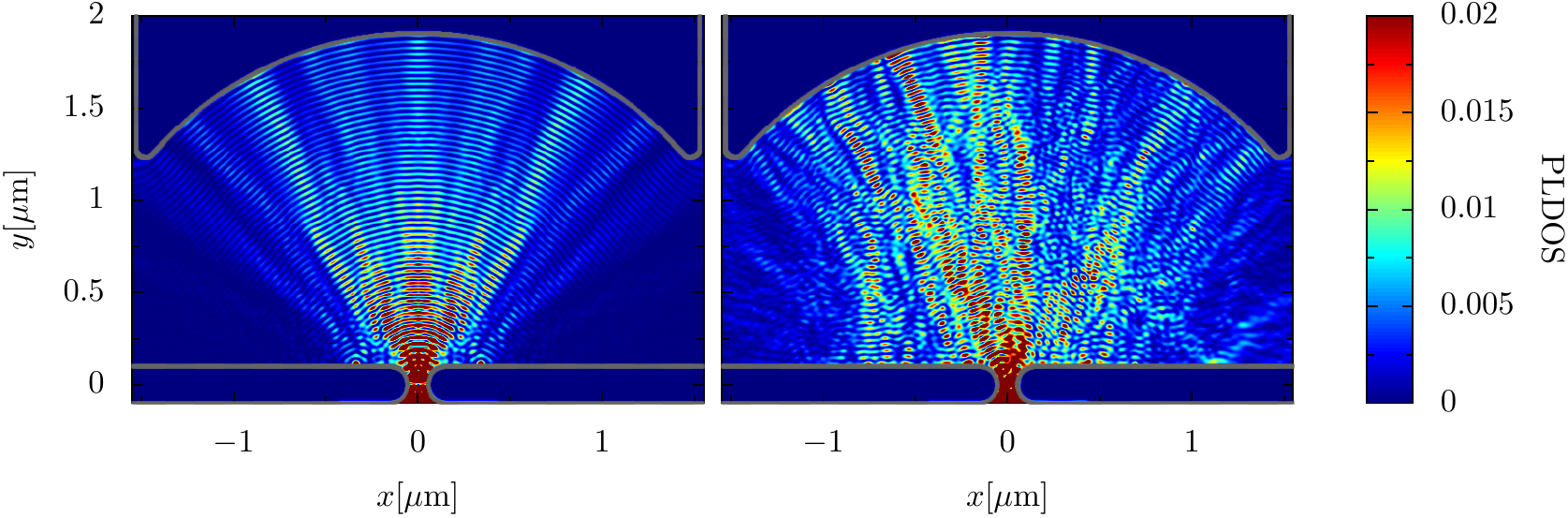}
\caption{Numerically calculated PLDOS at the Fermi energy corresponding to 	the scattering states entering the cavity through the QPC, for a depleting 	cavity gate. The left panel depicts the PLDOS for an ideal clean system, the right panel is for a sample with realistic disorder strength.}
\label{fig2}
\end{figure*}
The zero-temperature conductance fluctuations are suppressed at finite temperature.
In the clean case, the conductance exhibits regular Fabry--Perot like oscillations. 
In the presence of disorder, mesoscopic fluctuations on the scale of $\lambdaF/2$ survive like in the experiment. 
This is indicated by the horizontal scale bar of length $\lambdaF/2$. 
In our model, the shift of the edge of the depletion region is proportional to $\ln(\EF/\gamma_\mathrm{cav})$. 
The amplitude of the mesoscopic fluctuations calculated for the temperature of \SI{500}{\milli\kelvin} is a little bit larger than that in the experiment. 
This discrepancy could be resolved by further increasing the temperature for the calculation, in order to match the energy scale of the source-drain voltage used in the measurement.

The energy spacing of cavity modes can be experimentally determined using finite bias spectroscopy at different values of \Vqpc with \Vcav at a fixed value. 
For this purpose, the differential conductance $\mathrm{d}I/\mathrm{d}\VSD$ is measured as a function of \Vqpc and \VSD. 
The numerical derivative $\mathrm{d^2}I/\mathrm{d}\VSD^2$ is plotted in Fig.\,\ref{fig1}(c) in order to emphasize the cavity-mode related modulation of the differential conductance. 
The dashed lines indicate the diamond shaped region of zero differential conductance of the first QPC conductance plateau extracted directly from the differential conductance measurement. 
Marked by the white arrows, one recognizes a set of diagonal lines which are a manifestation of the cavity modes. 
The average-spacing of these modes is $\Ecav=\SI[separate-uncertainty = true]{317(14)}{\micro\electronvolt}$, which is in good agreement with the energy separation obtained for a one-dimensional resonator with the length of the cavity radius.\footnote{Note that the bias spectroscopy was done after a cooldown where the gates were prebiased. Compared to the cooldowns with grounded gates an offset in \Vqpc of \SI{0.2}{\volt} has to be taken into account.}

\section{Scattering states in the cavity}

The experimental results of Fig.\ \ref{fig1}(b), as well as the simulations shown in (d), reflect the general design idea of the cavity to form a regular mode pattern in analogy to an electromagnetic wave mirror cavity.\cite{Hersch1999,Roessler2015} 
In an ideal semicircular cavity, the energy eigenstates would be described by a radial and an angular quantum number. 
The large openings at the sides of the cavity strongly couple high angular momentum states to large 2DEG regions. 
In contrast, the lowest angular momentum modes are least affected by the openings and contribute to the local density of states with a strong modulation in energy. 
Subsequent peaks in the density of states correspond to a sequence of modes. 
That is, states with an increasing number of nodes in radial direction, giving rise to oscillations similar to those observed in Figs.\ \ref{fig1}(b) and (d).
In Fig.\ \ref{fig2}(a), we show as an illustration the simulation of the partial local density of states (PLDOS) for scattering states injected into the cavity through the QPC at the Fermi energy \EF in the absence of disorder.
It is characterized by a spatial modulation in radial direction with a period of half the Fermi wavelength. 
A much smoother variation of the PLDOS exists in azimuthal direction. 
Most of the PLDOS is concentrated close to the QPC reflecting the focusing effect of the cavity gate.

This view on the disorder-free cavity invoking the confinement induced coherent modes contrasts with the observation of branched flow in open two-dimensional electron gases,\cite{Topinka2001} an effect  relying on small-angle scattering. 
Similarly here, the energy spectrum of the cavity is significantly altered by the unavoidable disorder potential present even in this high-mobility system [see Fig.\ \ref{fig2}(b)].
The disorder significantly perturbs the PLDOS that attains higher values close to the split-gate near the left and right openings of the cavity, consistent with the disorder-induced increase of the conductance in the presence of the cavity shown in Fig.\ \ref{fig1}(d). 
The azimuthal PLDOS-pattern of Fig.~\ref{fig2}(a) is severely altered. 
However, the $\lambdaF/2$-modulation in radial direction survives, in agreement with the observed cavity-mode related modulation of the conductance with $\Vcav$ in Fig.~\ref{fig1}(b).

\begin{figure*}
\centering
\includegraphics[width=0.85\textwidth]{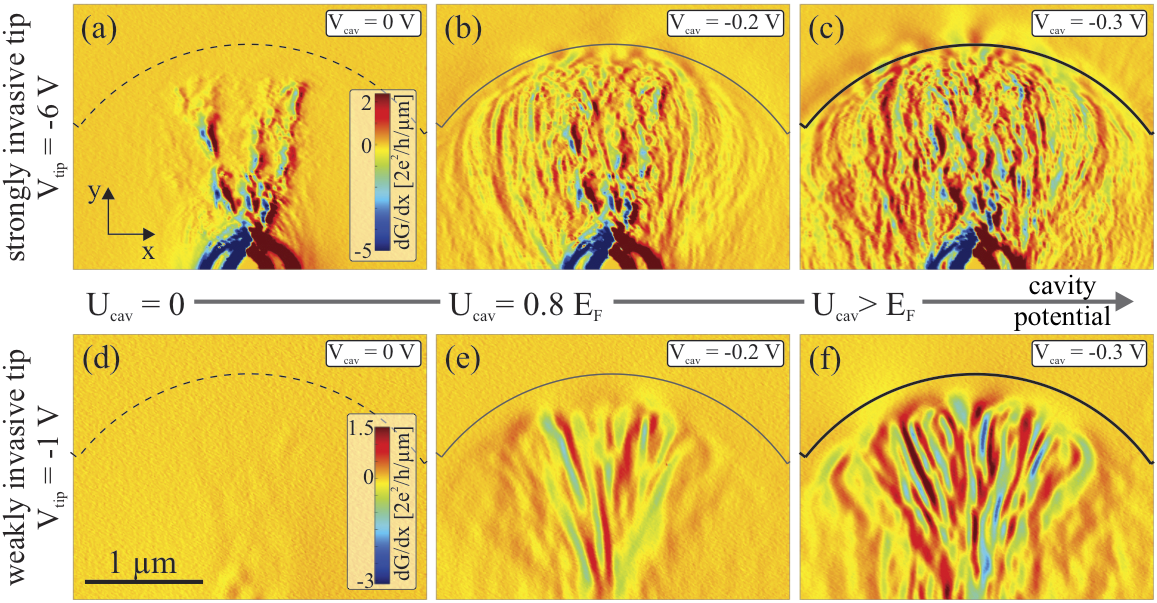}
\caption{Two cavity-gate voltage scan series showing the derivative $\mathrm{d}G/\mathrm{d}x$ of the conductance with respect to tip-position for	tip-voltage $\Vtip=\SI{-6}{\volt}$ in the upper panels (a)-(c), and for 	$\Vtip=\SI{-1}{\volt}$ in the lower panels (d)-(f).
The cavity voltage \Vcav is zero in the left panels (a) and (d). 
In the central panels (b) and (e), $\Vcav =\SI{-0.2}{\volt}$ does not deplete the 	2DEG under the gate, while the value $\Vcav =\SI{-0.3}{\volt}$ used in the right panels	(c) and (f) is beyond the depletion threshold.
The dashed gray, solid gray and solid black lines depict the outline of the cavity gate that becomes more important from left to right.}
\label{fig3}
\end{figure*}
\section{Effect of the cavity on scanning gate measurements}

We now turn to the discussion of scanning gate measurements, starting with a strongly invasive tip biased at $\Vtip=\SI{-6}{\volt}$, for which $U_\mathrm{t}>E_\mathrm{F}$. 
In Figs.\ \ref{fig3}(a--c) we plot the numerical derivative $\mathrm{d}G/\mathrm{d}x$ of the measured conductance $G(x,y)$ for different cavity gate voltages.
In Fig.\ \ref{fig3}(a) measured at $\Vcav=0$, we observe branched electron flow \cite{Topinka2001,heller2003branching,Kozikov2013,Steinacher2016} emanating from the QPC. 
The depletion disk induced in the 2DEG below the tip scatters electrons back through the QPC and reduces the conductance whenever placed above a region of high PLDOS. 
The branches cannot be seen when the tip scans above the cavity gate since the latter screens the \tippot.

For the scan in Fig.\ \ref{fig3}(b) we apply $\Vcav=\SI{-0.2}{\volt}\approx\VG{depl,cav}{}$ [cf.\ \ref{fig1}(b)] and just about avoid depleting the electron density below the cavity gate completely. 
We observe that the resulting potential barrier contributes significantly to electron-backscattering. 
The branch pattern disappears almost completely in a background of cavity-gate induced conductance modulations.
The branch pattern is no longer dominating the electron flow if the cavity is formed. 
This change of regime is discussed in detail in Ref.~\onlinecite{Steinacher2016}. 
It is due to multiple scattering of electrons between the \tippot and the cavity-potential induced by both the split-gate and the cavity gate.

Finally, Fig.\ \ref{fig3}(c) shows a scan at $\Vcav=\SI{-0.3}{\volt}$ where the 2DEG under the cavity gate is fully depleted. 
The conductance response is dispersed over the entire cavity area and the resulting conductance is modulated on the scale of $\lambdaF/2$. 
With the strongly invasive tip we induce conductance fluctuations resulting from an effective change of the sample geometry brought about by the scanning \tippot. However, from this measurement we cannot hope to obtain information about the cavity states present in the absence of the tip.

The set of measurements presented in Figs.\ \ref{fig3}(a)--(c), with the strong influence of the tip within the cavity region and the vanishing influence outside, shows experimentally that the cavity geometry indeed concentrates the local density of states within the cavity area, as suggested by the simulations in Fig.~\ref{fig2}(b).
In order to reduce the invasiveness of the tip, we now increase \Vtip and thereby decrease the tip-potential amplitude \Ut below \EF. 
We repeat the measurements shown in Fig.\ \ref{fig3}(a)--(c) at the same cavity gate voltages, but with $\Vtip=\SI{-1}{\volt}$, where $\Ut<\EF$. 
A smooth almost unstructured conductance map is obtained in Fig.\ \ref{fig3}(d) when the weakly-biased tip is scanned above the system with the grounded cavity-gate. 
This demonstrates the well-known fact that a weak potential perturbation cannot scatter electrons back through the QPC. 

As soon as the cavity-potential is raised to a sufficiently high amplitude 
[see Fig.\ \ref{fig3}(e)], a clear effect of the tip on the conductance is 
observed inside the area between cavity-gate and QPC. 
At $\Vcav=\SI{-0.3}{\volt}$ [see Fig.\ \ref{fig3}(f)] we measure rich spatial structures of the conductance with a pattern modulated in azimuthal direction emanating in fine strands from the QPC into the cavity area. 

Like for the case of a strongly invasive tip, rising the confinement of the system will cause an increasing concentration of the local density of states in the cavity at the Fermi energy. 
The weak \tippot perturbs this area of concentrated LDOS to significantly modify the conductance, attaining, for the strongest confinement, values that are comparable to those of the invasive tip [Fig.\ \ref{fig3}(f)]. 
In contrast to the strongly invasive regime the modification of backscattering induced by the tip is rather due to gentle electron lensing than to hard-wall reflection.

\section{Scanning gate measurements with varying tip-voltage}

\begin{figure}
\centering
\includegraphics[width=\linewidth]{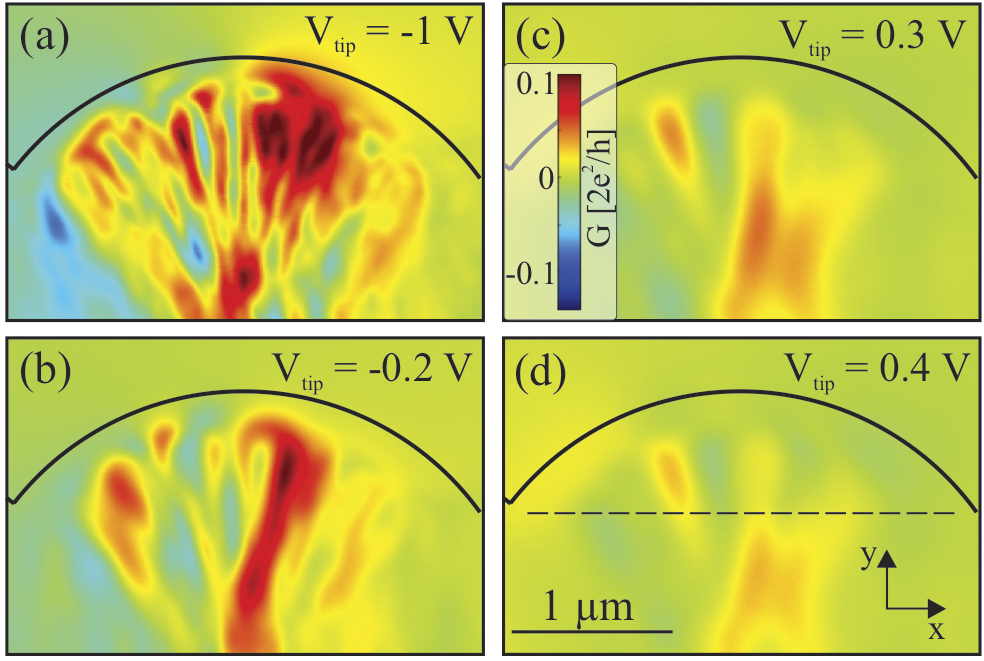}
\caption{Conductance change $\Delta G$ as a function of tip position $(x,y)$ for four different tip-voltages \Vtip for a cavity (indicated by solid lines) created by a voltage $\Vcav=\SI{-0.5}{V}$ which depletes the 2DEG underneath. 
The dashed horizontal line in panel (d) is the place chosen to study the characteristic spatial scale of the conductance map (see text).} 
\label{fig4}
\end{figure}

We measured a series of scans with varying \Vtip, some of which are displayed in Fig.\ \ref{fig4}. 
All plots show the change in conductance $\Delta G(x,y)=G(x,y)-G_0$ as a function of the tip position $(x,y)$ at a given tip-voltage \Vtip, where $G_0$ is the conductance of the system unperturbed by the tip (taken from the upper right corner of each scan). 
As stated in Sec.\ \ref{section3}, the quantum point contact voltage is kept at $\Vqpc=\SI{-0.5}{\volt}$, corresponding to the 3\textsuperscript{rd} mode. 
The cavity is set to a voltage of $\Vcav=\SI{-0.5}{\volt}<\VG{depl,cav}{}$ that produces complete depletion.

Figure \ref{fig4}(d) shows the scan with the value of \Vtip exhibiting the 
weakest effect on the conductance, which defines the least-invasive tip-voltage $V_\mathrm{li} = \SI{0.4}{V}$. 
A small decrease of \Vtip to \SI{0.3}{V} [Fig.\ \ref{fig4}(c)] yields a very similar spatial pattern as before, but with stronger contrast, as expected in the perturbative regime. \cite{Jalabert2010} 
We observe a smeared pattern for $\Vtip=\SI{-0.2}{V}$ [Fig.\ \ref{fig4}(b)] and a more structured one for $\Vtip=\SI{-1}{V}$ [Fig.\ \ref{fig4}(a)]. 

From measurements without the cavity, not shown here,\cite{steinacher2015scanning,Steinacher_thesis} we find that the \tippot starts to deplete the electron gas at a voltage of $V_\mathrm{depl,tip}=\SI{-5}{\volt}$ for which $\Ut=\EF$. Together with $V_\mathrm{li}$ this gives us the linear estimate $\Ut=\ut\times \EF$ for the amplitude of the \tippot as a function of \Vtip, where
\begin{equation}
\ut = \frac{\Vtip-V_\mathrm{li}}{V_\mathrm{depl,tip}-V_\mathrm{li}}.
\label{calibration}
\end{equation}
The values of \ut obtained from this estimate are given in Fig.\ \ref{fig4}, the Fermi energy in our sample is $\EF = \SI{5.3}{\milli\electronvolt}$. 

COMSOL simulations of the electrostatic problem, treating screening of the \tippot within the Thomas--Fermi approximation, confirm the linear dependence of \Ut on \Vtip as long as $\Ut\leq\EF$. 
The potential profile is found to be approximately Lorentzian in this regime, \cite{Eriksson1996,Pala2008,steinacher2015scanning} while for $\Ut>\EF$ a Gaussian profile better describes the tail potential. \cite{Steinacher_thesis}

\begin{figure}
\centering
\includegraphics[width=\linewidth]{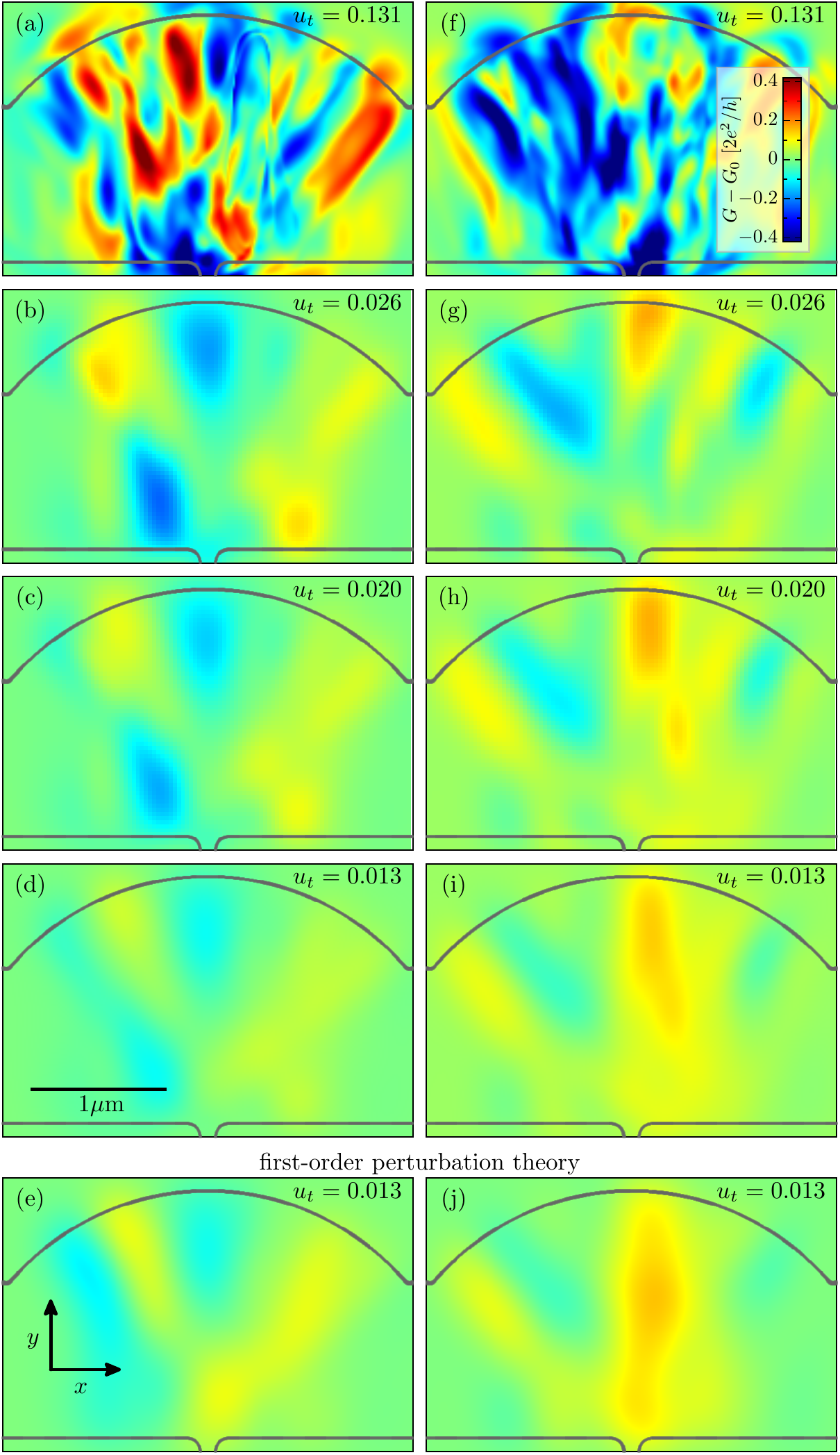}
\caption{Numerically calculated tip-induced conductance corrections for two different disorder configurations (left and right panels) as a function of tip	position $(x,y)$, for the QPC operating on the third plateau and in the presence of a strong reflector cavity potential (indicated by solid lines). 
The upper panels (a) and (f) correspond to a large tip strength, while the panels (b)--(d) and (g)--(i) depict the conductance change due to a weakly invasive tip of 	decreasing strength. 
The lowest panels (e) and (j) show the prediction of first-order perturbation theory (\ref{eq:first_order}) that describes well the effect of a weak tip.} 
	\label{fig5}
\end{figure}
Qualitative behavior of the conductance similar to that observed experimentally in Fig.\ \ref{fig4} is obtained from the numerical simulations shown in Fig.\ \ref{fig5}. 
The \tippot was modeled as a Gaussian with peak amplitude $\Ut=\ut\times\EF$, centered at $\mathbf{r}_\mathrm{tip}$, and a width parameter $\sigma=\SI{177}{\nano\meter}$. 
The choice of a Gaussian profile for the simulations is made for simplicity, and in order to eliminate the effects of long-range tails in the tip potential.

A quantitative agreement between details of the spatial conductance pattern in experiment and theory is not expected for several reasons. 
First, the realization of the specific disorder potential in the simulation cannot agree with the experimental one, although the statistical parameters (characteristic modulation amplitude and correlation length) are chosen to reproduce the experimental mobility.
Second, the simulations are restricted to zero temperature, in contrast to the experiment.
Third, it is hard to exactly quantify the width of the experimental \tippot and compare it to the simulations.

\begin{figure}
	\centering
	\includegraphics[width=\linewidth]{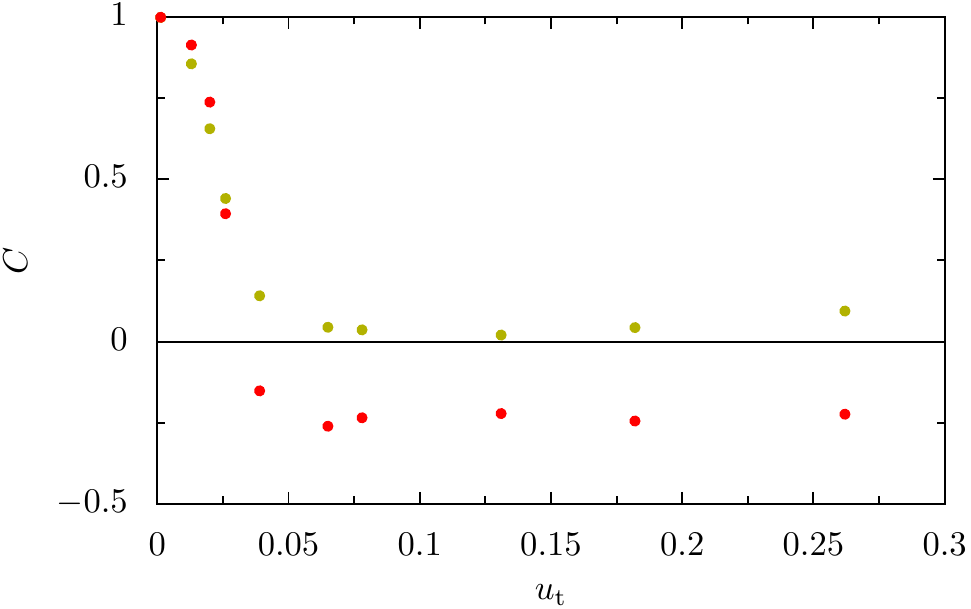}
	\caption{The correlation $C$ of the numerically calculated conductance correction $G(x,y)-G_0$ in the cavity area with the prediction of first-order perturbation theory (\ref{eq:first_order}), as a function of the tip strength $u_\mathrm{t}$. Olive and red dots are results for the realizations used in the left and right column in Fig.\ \ref{fig5}, respectively.} 
	\label{fig6}
\end{figure}
The results of the simulations are shown in Fig.\ \ref{fig5} for two different impurity configurations (left and right column of the plot).
They reproduce the experimental tendency to evolve from a structured 
 [Fig.\ \ref{fig5}(a) and (f)] to a smooth pattern when the tip strength is 
reduced (in absolute value) taking us from the invasive to the non-invasive regime [Fig.\ \ref{fig5}(d) and (i)]. 

Comparing the experimental result of Fig.\ \ref{fig4}(b) with the calculated ones of Figs.\ \ref{fig5}(a) and (f) with comparable \ut one recognizes two differences. 
On one hand, there is fine-structure on a shorter length scale in the calculations than in the experiment. 
On the other hand, the amplitude of the conductance modulations is roughly four times larger in the simulated data than in the experiment. 
Most likely, the differences would both disappear, if a finite temperature were introduced in the calculation, with the obvious effect of smearing spatial structure and reducing the modulation amplitude. 
However, generating a complete scan at finite temperature is computationally very costly.
The agreement obtained between the SGM measurements and the calculated scans is limited for the reasons given above, but it still supports the modelling used to interpret the experiments. 

\section{Perturbative regime}

The series of simulations shown in Fig.\ \ref{fig5}(a--d) and (f--i) give us the opportunity to connect to previous exact theoretical results treating the tip-potential perturbatively. 
The goal is to find by comparison the range of validity in \ut for the perturbative treatment in the present realistic scenario.
To lowest order in the tip potential, the conductance change induced by the tip is \cite{Jalabert2010,Gorini2013}
\begin{equation}\label{eq:first_order}
	G^{(1)} =  - \frac{8 \pi e^2}{h}	\mathrm{Im}\left[\mathrm{Tr}\left\{ r^\dagger t' \mathcal{V}^{2,1}\right\}\right] \, ,
\end{equation}
where $r$ and $t'$ are the reflection (from the left) and the transmission (from the right) matrices, respectively. The entries of $\mathcal{V}^{2,1}$ are obtained as the matrix elements
\begin{equation}\label{eq:matrix_element}
\mathcal{V}^{2,1}_{a,a'}(\mathbf{r}_\mathrm{tip})=\int \mathrm{d}^2r \, \psi^{(2)*}_a(\mathbf{r})U_\mathrm{tip}(\mathbf{r},\mathbf{r}_\mathrm{tip})\psi^{(1)}_{a'}(\mathbf{r})
\end{equation}
of the tip potential $U_\mathrm{tip}(\mathbf{r},\mathbf{r}_\mathrm{tip})$ between scattering states impinging from leads 1 and 2.
The conductance correction obtained from Eq.\ \eqref{eq:first_order} by using the unperturbed scattering wave functions, the scattering amplitudes and the tip potential [Fig.\ \ref{fig5}(e) and (j)] reproduce with a good accuracy the corresponding conductance corrections of Fig.\ \ref{fig5}(d) and (i), showing that for those system parameters we are in the perturbative regime.

In order to determine the extension of the regime of validity of first-order perturbation theory, we evaluate the correlation coefficient between the full calculated SGM response and the first-order prediction \eqref{eq:first_order}.
The result for the two disorder realizations of Fig.\ \ref{fig5} is shown in Fig.\ \ref{fig6}, indicating that the first-order conductance correction provides a good description of the SGM response up to a tip strength of about $u_\mathrm{t}\approx 0.025$. With the relation \eqref{calibration}, that value corresponds to tip voltages of 
about \SI{0.135}{\volt} around the least invasive value $V_\mathrm{li}$.  

At the same time, the experiment yields significant conductance changes in  the regime where the first-order term \eqref{eq:first_order} is the dominant conductance correction [see Fig.\ \ref{fig4}(c,d)].
Within this regime, a change in tip voltage leads to a change of the global prefactor of the SGM response, without altering the spatial pattern. 
We thus confirm that our special purpose setup is able, for the first time, to access the weakly-invasive regime, where the interpretation of the scanning gate measurements is unambiguously given by non-local properties of the unperturbed system.\cite{Jalabert2010,Ly2017} 
In the next section we will investigate in how far we can experimentally exploit this theoretical finding.

\section{Characteristic scales and related SGM regimes}
\begin{figure*}
\centering
\includegraphics[width=0.7\linewidth]{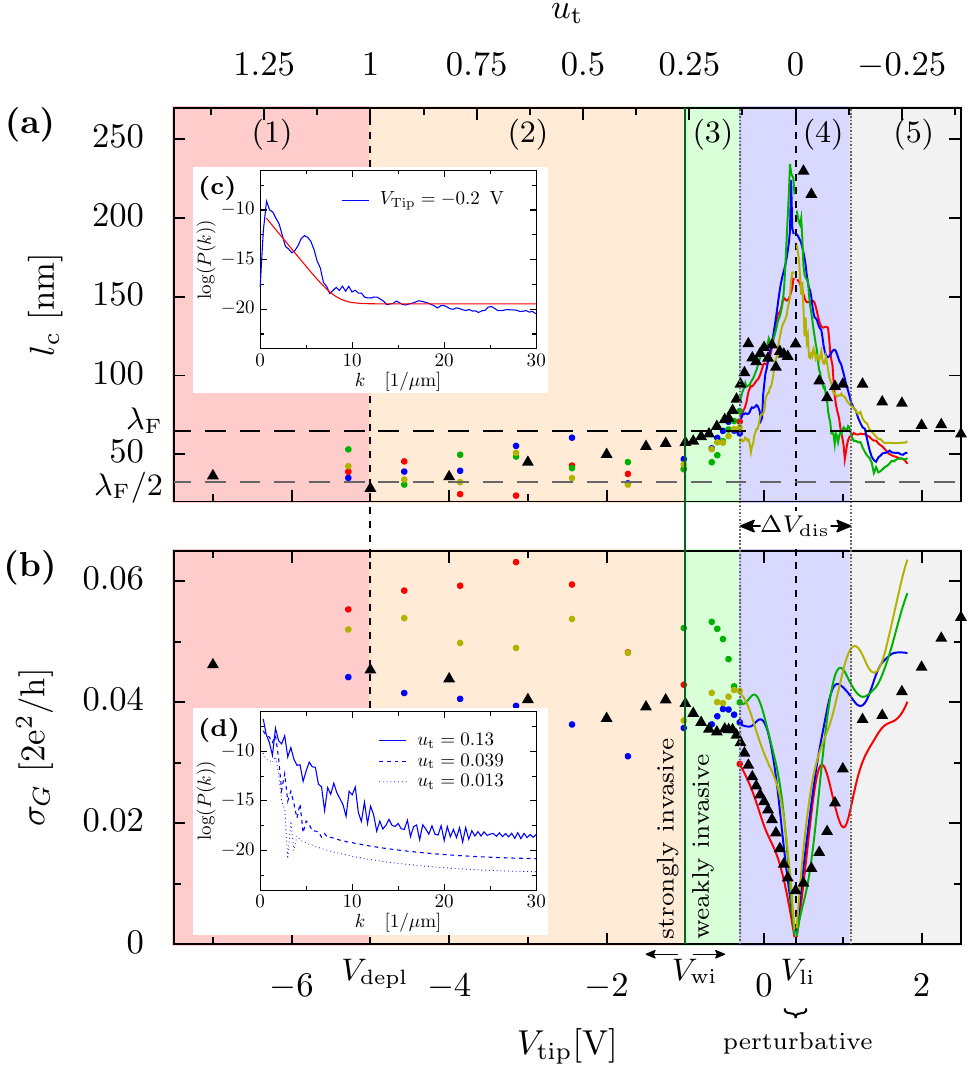}
\caption{(a) Black triangles: characteristic length scale \lc as a function of \Vtip (lower axis), extracted from the experimental SGM response $P(k)$ as described in the text and shown in the insets for two values of the tip voltage. 
Colored lines and points: \lc extracted from the numerically generated SGM response at $T=\SI{500}{\milli\kelvin}$ as a function of the corresponding tip strength \ut (upper axis) for four different impurity realizations.
The dashed horizontal lines show intrinsic length scales of the system: the Fermi wave length $\lambdaF$, and its half $\lambdaF/2$. 
(b) The standard deviation $\sigma_G$ of the conductance data as a function of the tip-voltage from the experimental measurements (black triangles) and $\sigma_G/2$ for the same four disorder configurations used in (a) (colored lines and points). 
Three characteristic voltages are marked: \VG{li}{} is the least-invasive 	voltage; \VG{depl}{} the depletion voltage of the tip; \VG{wi}{} identifies the 	transition from the weakly to the strongly-invasive \Vtip. 
The disorder-broadening $\Delta \VG{dis}{}$ is indicated by dotted black 	lines. 
The conversion between the theoretical tip strength parameter \ut and \Vtip has been assumed to be linear with parameters chosen such that the experimental data in the weakly invasive regime are similar to the average of the numerically obtained ones.
(c) The power spectrum (blue line) of the tip-induced conductance change 
measured at 	$\Vtip=\SI{-0.2}{V}$ along the dashed line in Fig.\ \ref{fig4}(d). 
The red line is the fit (see text) to extract the characteristic value $k_\mathrm{c}$.
(d) The power spectrum found in the theoretical simulation for the disorder realization that is represented by blue lines and points in (a) and (b).
Dotted and dashed lines are for different weak tip strengths, the solid line is for a moderately strong tip.}
\label{fig7}
\end{figure*}

The previous observations call for a quantitative analysis of the extent of the perturbative regime (in $\Vtip-V_\mathrm{li}$) and the change in the spatial resolution of the conductance maps when going from Fig.\ \ref{fig4}(a) to (c) [and from Figs.\ \ref{fig5}(a) and (d) to \ref{fig5}(b) and (e)]. 
For this purpose we heuristically define a characteristic length scale \lc for the conductance modulations seen in the SGM images. 
For every scan at a fixed tip-voltage, we extract the conductance along the dashed line shown in Fig.\ \ref{fig4}(d) at $y=y_\mathrm{s}$. 
The power spectral density $P(k)=|\widetilde{\delta G}(k)|^2$ of a one-dimensional fast Fourier transform is calculated for the measured function $\delta G(x,y_\mathrm{s})=G(x,y_\mathrm{s})-\overline{G}_{y_\mathrm{s}}$, where $\overline{G}_{y_\mathrm{s}}$ is the average conductance along the line $y=y_\mathrm{s}$. 
Figure \ref{fig7}(c) shows an example of such a power spectrum for $\Vtip=\SI{-0.2}{V}$.

The experimentally obtained power spectra do not contain clearly distinguishable peaks, but they show an approximately exponential decay at small $k$ followed by a nearly constant background at large $k$, with the transition between the two regimes at the characteristic scale $k=k_\mathrm{c}$. 
The value $k_\mathrm{c}$ allows us to define a characteristic length scale observed in the SGM measurements through $\lc(\Vtip)=k_\mathrm{c}^{-1}(\Vtip)$. 

A practical way of determining $k_\mathrm{c}$ is by fitting $P(k)$ with the function $f(k) = a\, e^{-b\,k} + c$, and identifying the crossover point as that of the largest curvature, i.e. $k_\mathrm{c}=b^{-1}\ln(a/c)$.
In Fig.\ \ref{fig7}(a), the black triangles show the experimentally determined values of $l_\mathrm{c}$ for values of \Vtip ranging from the strongly invasive to the perturbative regime.
Note that point-to-point variations of the measured \lc-values can arise from poor self averaging and little statistics in the case of \Vtip close to \VG{li}{}.
The characteristic scale $l_\mathrm{c}$ is seen to be maximum and almost constant around \VG{li}{} taking the value \SI{120}{\nano\meter}. 
For more negative tip voltages $l_\mathrm{c}$ decays monotonically, leveling off at about $\lambdaF/2$. For more positive tip voltages a similar, but less clear trend is observed.
Here the situation is quantitatively different, because in regime \rv the positive tip-voltage may induce a screening charge in the doping plane \cite{Rossler2010} in our heterostructure.

In addition to $l_\mathrm{c}$, we extract the standard deviation $\sigma_G$ of the conductance on the cross-sectional lines $G(x,y_\mathrm{s})$ to quantify the amplitude of the measured conductance modulation. 
The black triangles in Fig.\ \ref{fig7}(b) represent $\sigma_G$ as a function of tip-voltage. 
It is minimum at \VG{li}{} and increases almost linearly for more negative \Vtip. At tip-voltages $\Vtip<\SI{-1}{V}$ it levels off and shows only a weak increase with decreasing \Vtip. 
Figure \ref{fig7} thus shows quantities related to the length scale and amplitude of the observed conductance modulation present in the scans of Fig.\ \ref{fig4}(a)--(c).

In Fig.\ \ref{fig7} we empirically identify five distinct regimes labeled \ri-\rv. 
The boundaries between them will be explained in the following. In regime \ri, the tip produces a well-defined depletion disk within the 2DEG, such that branched electron flow would be observed at $\Vcav=0$ [cf. Fig.\ \ref{fig3}(a)]. 
When the cavity is formed, the conductance map shows modulations limited by $\lc \approx \lambdaF/2$ and with an amplitude $\sigma_G$ essentially independent of $\Vtip$ [cf. Fig.\ \ref{fig3}(b,c)]. 
In regime \rii, marked by the depletion voltage $\VG{depl}{}=\SI{-5}{V}$ for which the tip no longer induces a depletion of the 2DEG, the branches are not observed at $\Vcav=0$. 
The characteristic length scale \lc observed in the cavity weakly increases with \Vtip and attains values of the order of $\lambdaF$. 
In this regime $\sigma_G$ decreases slowly. 
Regime \riii is characterized by a strong increase of \lc within a small tip voltage range, while $\sigma_G$ still decreases very little [cf. Figs.\ \ref{fig3}(d--f) and Figs.\ \ref{fig4}(a,b)]. 
In regime \riv $\sigma_G$ reduces strongly until a minimum modulation amplitude is reached at the least invasive tip voltage $\VG{li}{}=\SI{0.4}{V}$ [cf. Fig.\ \ref{fig4}(c,d)]. 
At this voltage the sign of the \tippot changes. 
Beyond this voltage $\sigma_G$ increases symmetrically. 
In the same regime \lc assumes large values of more than a hundred nanometers. In regime \rv the length scale of the observed fluctuations decreases significantly while $\sigma_G$ continues to increase.

In order to interpret the physical significance of the five regimes from strong to weak tip strength we convert \Vtip into \ut using Eq.\ \eqref{calibration}.
The corresponding scale for \ut is found in Fig.\ \ref{fig7} as the upper horizontal axis of (a).
The characteristic amplitude of the disorder potential estimated from a model calculation \cite{Steinacher2016} results in 0.13\,\EF. 
This value agrees well with the width of regime \riv. 
In this regime the \tippot is comparable in strength to the disorder potential. 
The correlation length of the disorder potential $\ell_\mathrm{dis}$, extracted using the autocorrelation of our modeled disorder is approximately \SI{100}{\nano\meter}, and therefore roughly of the order of the maximum values that \lc attains.
However, also other length scales of the experiment, such as the tip--2DEG separation and the tip radius are of a similar magnitude. Further theoretical considerations are needed to clarify the significance of regime \riv. 
We therefore investigate the characteristic length \lc and $\sigma_G$ of the numerically calculated SGM response below. 
We estimate the SGM response close to the least-invasive condition using perturbative expressions in Sec.\ \ref{sec:interpretation}, and we present a perturbative argument based on the modulation of the density of states in the Appendix.

In order to simulate the experimental data for $\lc$ and $\sigma_\mathrm{G}$, we numerically calculate the tip-induced conductance change at a temperature of \SI{500}{\milli\kelvin} and perform the Fourier transform along the dashed line of Fig.\ \ref{fig4}(d) as in the experiment.
Shown in Fig.\ \ref{fig7}(d), the power spectrum corresponding to two weak tips ($\ut=0.013$ and $0.039$), and a moderately strong one ($\ut=0.13$) exhibit an exponential decay followed by an algebraic tail.
The tail arises from the finite interval on which the Fourier transform is performed, while the signature of the lattice used in the discretization of the system appears only at very large $k$ values that lie outside range of the figure. 

Using the same data analysis as in the experiment allows us to extract the crossover values $k_\mathrm{c}$ marking the end of the exponential regime, and the lengths \lc that can be compared to the experimental values.
As shown in Fig.\ \ref{fig7}, the behavior of \lc and $\sigma_G$ extracted from the numerically obtained SGM response for four different disorder realizations (colored dots) is similar to the experimental observation. 
The dispersion in the maximum values of \lc among the four samples is due to the different impurity configurations used in the simulations.
The quantitative discrepancy in the values of $\sigma_\mathrm{G}$ is most likely related to the discrepancy between the finite electronic temperature in the experiment and the zero temperature simulation.

The observed similarity of the results for several different disorder realizations is strong support for the generic character of our numerical simulations and of the experimental data that have been obtained within a single sample.

\section{Interpretation within the perturbative approach}
\label{sec:interpretation}

The reduction of \lc and the increase of $\sigma_G$ away from the least-invasive tip-voltage, found in the experiments and in the simulations shown in Fig.\ \ref{fig7}, can be understood within the perturbative theory.
For this analysis we assume that the \tippot is a function of fixed shape with a linear dependence on tip strength \ut, reading 
\begin{equation}\label{eq:tippot}
U_\mathrm{tip}(\mathbf{r},\mathbf{r}_\mathrm{tip})=\EF\times \ut v(\mathbf{r}-\mathbf{r}_\mathrm{tip}).
\end{equation}

In the limit of very weak \ut the average value $\overline{G}_{y_\mathrm{s}}$ along the scanning line approaches the unperturbed conductance $G_0$. Thus, $\delta G(x,y_\mathrm{s})=\Delta G(x,y_\mathrm{s})$ and 
\begin{equation}\label{eq:ft-perturbative}
P(k)=\left|\sum_n \widetilde{G^{(n)}}(k)\right|^2\, ,
\end{equation}
where $\widetilde{G^{(n)}}(k)$ is the Fourier transform of the $n$\textsuperscript{th} order conductance correction proportional to $\ut^n$.

Staying within first order, by using Eqs.\ \eqref{eq:first_order} and \eqref{eq:matrix_element}, leads to the proportionality of $\sigma_G$ with $|\ut|$. Such a behavior is indeed observed in Fig.\ \ref{fig7}(b). The first order correction sets the maximum value attained by \lc at the least-invasive tip-voltage. Taking the scanning interval as infinite,  $\widetilde{G^{(1)}}(k)$ can be written as
\begin{eqnarray}
&&\widetilde{G^{(1)}}(k)=-\frac{8\pi e^2}{h}\ut\EF \, \mathrm{Im}\bigg[ \int \mathrm{d}y \, \tilde{v}(-k,y-y_\mathrm{s})\times \nonumber \\ 
\label{eq:FT}&&\quad\sum_{a,a',a''} \! r_{a,a'}^\dagger t_{a',a''}^\prime \! \int\! \mathrm{d}x\, e^{-2\pi i kx} \psi^{(2)*}_{a''}(x,y)\psi^{(1)}_{a}(x,y) \bigg]
\end{eqnarray}
in terms of the one-dimensional Fourier transform of the tip potential $\tilde{v}(-k,y-y_\mathrm{s})$ and the wave-function products (last integral in \eqref{eq:FT}).

The low-$k$ sector of the power spectrum $P(k)$, where $k\lesssim 1/w$, is therefore dominated by the decay of the Fourier-transform of the \tippot, modulated by the azimuthal variations of the scattering wave function product.
In the perturbative regime, the slope $-b$ on the logarithmic scale follows therefore from the characteristic length scale of $v(\mathbf{r})$ and may be used to estimate $w$ from the measurement. In particular, if the tip profile  $v(\mathbf{r})$ has a Lorentzian shape of width $w$, we have $b=w$.
 
The high-$k$ sector of $P(k)$ results from the spatial variations of the scattering states on the scale $\lambdaF$, as well as from corrections to the Fourier transform caused by the finite spatial length of the analyzed traces in $x$-direction.

Understanding the reduction of \lc away from the least-invasive tip-voltage necessitates to consider higher-order terms of the perturbative expansion \eqref{eq:ft-perturbative}.
The $n$\textsuperscript{th} order correction contains products of $n$ matrix elements similar to the one of Eq.\ \eqref{eq:matrix_element} \cite{Jalabert2010,Gorini2013}, such that the spatial dependence of $G^{(n)}$ contains the $n$\textsuperscript{th} power of the tip shape \eqref{eq:tippot}. 
Thus, higher-order conductance corrections are expected to reveal sharper and sharper structures. 
The corresponding Fourier transforms $\widetilde{G^{(n)}}(k)$ then have a slower and slower decay in the low-$k$ sector. 
Their contribution in \eqref{eq:ft-perturbative} is increasing with \ut, explaining the increase of $k_\mathrm{c}$ and the decrease of \lc with growing tip strength until the characteristic length scale of the wave-functions $\lambda_\mathrm{F}/2$ is reached, as observed in Fig.\ \ref{fig7}.
The previous argument for the decrease of \lc away from $\ut=0$ remains qualitative, since the quantification of the higher-order conductance corrections becomes progressively more complicated.
Nevertheless, the perturbative theory allows to explain the impressive resolution of SGM experiments using strong tips that is enhanced beyond the size of the tip potential, and the absence of such an enhancement in the weakly invasive regime.  

\section{Conclusions}

In this work, we presented scanning gate measurements on a tunable cavity, demonstrating that the obtained results with a weak tip-potential are in great contrast to experiments with a strongly-invasive tip. 
The confinement of our cavity geometry with a weakly perturbing tip gives rise to a regime where the tip potential only gives the amplitude of the SGM map, without altering its spatial distribution.
We could then access and quantify the extent of the perturbative regime, where the SGM response is unequivocally related to properties of the unperturbed system.

The good agreement found between the SGM measurements and the quantum simulations allowed us to identify the imaging mechanisms according to the strength of the tip-potential. 
We found that the size of the \tippot plays a major role for the spatial resolution, and as a consequence, couples the tip's invasiveness and its resolution. 
These results lead to a new understanding of weakly-invasive scanning gate imaging. 

The finite tip size needs to be considered when it comes to the interpretation of scanning gate results. 
In the weakly invasive regime, the spatial resolution is limited by the tip geometry, such that a direct connection between the observed SGM response (Figs.\ \ref{fig4} and \ref{fig5}) with the LDOS (Fig.\ \ref{fig2}) cannot be made. 

The development of sharper tip-potentials, difficult to realize experimentally, would be of great help to gain in resolution for the SGM technique in the weakly invasive regime. Only in a recently developed SGM setup for cold atomic gases\cite{Haeusler2017} a tip size comparable to the Fermi wavelength has been achieved.

Increasing the tip-strength and entering into the invasive regime results in sharper spatial structures, but the interpretation of the SGM measurements in a confined geometry is not as clear-cut as in the non-invasive case.
The increase of the spatial resolution with tip-voltage strength has been characterized in terms of a length parameter whose behavior can be understood from perturbation theory.  

The connection of SGM measurements with the LDOS of the unperturbed system has been actively pursued in recent experimental and theoretical studies, and our work contributes to such a quest.
For one-dimensional systems, when the SGM setup is modeled by a tight-binding system, a perturbative approach was able to link the first-order conductance change produced by a $\delta$-tip with the Hilbert transform of the local density of states.\cite{Pala2008,gasparian1996} 
In the case of perfect transmission, i.\ e.\ a QPC with an exact conductance quantization \cite{Gorini2013,Ly2017} or a quantum dot operating close to a resonance,\cite{kolasinski2013} the first-order conductance correction in the tip strength (\ref{eq:first_order}) vanishes, and the second-order correction dominates the SGM response, which turns out to be proportional to the square of the LDOS. 
For quantum dots outside the resonance condition, which is the case studied in this work, the first-order correction (\ref{eq:first_order}) that we have shown to dominate the SGM response for weak tips, has no evident correspondence with the LDOS.\cite{kolasinski2013, Ly2017}  

\begin{acknowledgments}	

We thank Oded Zilberberg, Michael Ferguson, Jana Pijnenburg, Clemens R{\"o}ssler, Beat Br{\"a}m, and Ousmane Ly for helpful discussions. We acknowledge financial support from the Swiss National Science Foundation, ETH Z{\"u}rich, and the French National Research Agency ANR through projects ANR-11-LABX-0058\_NIE (Labex NIE) within ANR-10-IDEX-0002-02 and ANR-14-CE36-0007-01 (SGM-Bal).
\end{acknowledgments}

\appendix*
\section{A density of states approach}

In this appendix we provide an alternative approach to that of Sec.\ \ref{sec:interpretation} allowing us to qualitatively understand the decrease of the characteristic length \lc away from the least-invasive condition $\ut=0$.
We will then assume that the conductance changes of the structure are governed by the mesoscopic density of states fluctuations in the cavity region at the characteristic energy scale \Ecav witnessed by the measurements shown in Fig.\ \ref{fig1}(c). The central idea of this approach is that the conductance on a QPC-plateau is modulated by the density of states of the cavity at the Fermi energy. Scanning the \tippot at position $\mathbf{r}_\mathrm{tip}=(x_\mathrm{tip},y_\mathrm{tip})$ into the cavity region will scramble the quantum states in the cavity, which changes the cavity density of states at the Fermi energy and thereby modulates the conductance $G(\mathbf{r}_\mathrm{tip})$.

We estimate this effect by arguing that the \tippot in Eq.\ \eqref{eq:tippot} will lift the conduction band bottom within the cavity on average by
\[ \Delta E_\mathrm{c}(\mathbf{r}_\mathrm{tip}) \approx \frac{\ut\EF}{\Acav}
\int_\mathrm{cav}\mathrm{d}^2r \, v(\mathbf{r}-\mathbf{r}_\mathrm{tip}), \]
where \Acav is the cavity area. The density of states modulation occurs on the scale $\Delta E_\mathrm{c}(\mathbf{r}_\mathrm{tip})\sim\Ecav$, which translates into a correlation distance $\Delta\mathbf{r}_\mathrm{tip}$ determined by
\[ \Ecav \approx \Delta \mathbf{r}_\mathrm{tip} \frac{\ut\EF}{\Acav} 
\frac{\mathrm{d}}{\mathrm{d}\mathbf{r}_\mathrm{tip}}
\int_\mathrm{cav}\mathrm{d}^2r \, v(\mathbf{r}-\mathbf{r}_\mathrm{tip})\, . \]
For the derivative of the integral over the tip shape with respect to the tip position, it is crucial to realize that the integral is limited to the finite area of the cavity. 
The strongest dependence on tip position occurs when the maximum of the tip potential crosses the edge of the cavity. For a simple straight edge parallel to the $y$-axis and a displacement of the tip in $x$-direction the value of the derivative is given by the $y$-integral over $v(\mathbf{r}-\mathbf{r}_\mathrm{tip})=v(x-x_\mathrm{tip},y-y_\mathrm{tip})$. 
The derivative takes its maximum value when this integral is maximal. For circular symmetric tip shapes, we thus have
\begin{equation}
\Ecav \lesssim \Delta x_\mathrm{tip} \frac{\ut\EF}{\Acav} \int \mathrm{d}y \, v(0,y-y_\mathrm{tip})\, .
\end{equation}
We see that the best conceivable resolution is determined by the integral over a cut through the tip-potential. 
In the case of a Lorentzian \tippot and full width at half maximum $2w$, $v(\mathbf{r})=w^2/(r^2+w^2)$, one estimates
\begin{equation}
\frac{\Delta x_\mathrm{tip}}{w} \gtrsim \frac{1}{\ut}\frac{\Ecav }{\EF}\times\frac{\Acav}{\pi w^2}
\, .
\end{equation}
We see that this approach predicts a decrease of the correlation length with increasing strength of the \tippot in qualitative agreement with the experimental findings in regions \ri-\riii in Fig.\ \ref{fig7}(a). 
It also predicts that the correlation length can be significantly smaller than the width $w$ of the \tippot, if $\Ecav\ll\EF$.

Identifying $\Delta x_\mathrm{tip}$ with \lc we recognize the decrease of this characteristic scale away from the least-invasive tip-voltage.
However, the previous argument is not able to describe the limit of very small or very large \ut.

\bibliography{cavityBIB}

\end{document}